\pdfoutput=1
\documentclass[10pt, conference, compsocconf]{IEEEtran}

\ifCLASSINFOpdf
\else
\fi

\IEEEoverridecommandlockouts

\usepackage{tikz}
\usepackage{textcomp}
\usepackage{hyperref}
\usepackage{lipsum}

\newcommand\copyrighttext{%
  \footnotesize \textcopyright 2021 IEEE. Personal use of this material is permitted.
  Permission from IEEE must be obtained for all other uses, in any current or future
  media, including reprinting/republishing this material for advertising or promotional
  purposes, creating new collective works, for resale or redistribution to servers or
  lists, or reuse of any copyrighted component of this work in other works.
  }
  
\newcommand\copyrightnotice{%
~\\\\\\\\[3pt]
\copyrighttext
}

\usepackage{cite}
\usepackage{amsmath,amssymb,amsfonts}
\usepackage{algorithmic}
\usepackage{graphicx}
\usepackage{textcomp}
\usepackage{xcolor}
\def\BibTeX{{\rm B\kern-.05em{\sc i\kern-.025em b}\kern-.08em
    T\kern-.1667em\lower.7ex\hbox{E}\kern-.125emX}}

\usepackage{kotex} %
\usepackage[linesnumbered,ruled]{algorithm2e}

\usepackage{multirow} %

\usepackage{cancel,soul}
\usepackage[normalem]{ulem}

\usepackage{booktabs} 
\usepackage{caption}

\usepackage{pgfplotstable,filecontents}
\pgfplotsset{compat=1.9}%

\begin{document}
\bstctlcite{BSTcontrol}

\title{Intent-based Product Collections for E-commerce \\using Pretrained Language Models}

\author{\IEEEauthorblockN{
Hiun Kim\IEEEauthorrefmark{1},
Jisu Jeong\IEEEauthorrefmark{1}\IEEEauthorrefmark{2},
Kyung-Min Kim\IEEEauthorrefmark{1}\IEEEauthorrefmark{2},
Dongjun Lee\IEEEauthorrefmark{3},
Hyun Dong Lee\IEEEauthorrefmark{1},\\
Dongpil Seo\IEEEauthorrefmark{1},
Jeeseung Han\IEEEauthorrefmark{1},
Dong Wook Park\IEEEauthorrefmark{1},
Ji Ae Heo\IEEEauthorrefmark{1},
Rak Yeong Kim\IEEEauthorrefmark{1}}
\\
\IEEEauthorblockN{
\IEEEauthorrefmark{1}NAVER CLOVA\,\,\,\,\,\,
\IEEEauthorrefmark{2}NAVER AI LAB\,\,\,\,\,\,
\IEEEauthorrefmark{3}LBox Co., Ltd.\\\\
}}

\maketitle
\makeatletter
\def\ps@IEEEtitlepagestyle{
  \def\@oddfoot{\mycopyrightnotice}
  \def\@evenfoot{}
}
\def\mycopyrightnotice{
  {\footnotesize
  \begin{minipage}{\textwidth}
    \copyrightnotice
  \end{minipage}
  }
}

\begin{abstract}
Building a shopping product collection has been primarily a human job.
With the manual efforts of craftsmanship, experts collect related but diverse products with common shopping intent that are effective when displayed together, e.g., backpacks, laptop bags, and messenger bags for freshman bag gifts. %
Automatically constructing a collection requires an ML system to learn a complex relationship between the customer's intent and the product's attributes.
However, there have been challenging points, such as 
1) long and complicated intent sentences, 
2) rich and diverse product attributes,
and 3) a huge semantic gap between them, 
making the problem difficult. 
In this paper, we use a pretrained language model (PLM) that leverages textual attributes of web-scale products to make intent-based product collections. %
Specifically, we train a BERT with triplet loss by setting an intent sentence to an anchor and corresponding products to positive examples.
Also, we improve the performance of the model by search-based negative sampling and category-wise positive pair augmentation.
Our model significantly outperforms the search-based baseline model for intent-based product matching in offline evaluations.
Furthermore, online experimental results on our e-commerce platform show that the PLM-based method can construct collections of products with increased CTR, CVR, and order-diversity compared to expert-crafted collections.

\end{abstract}


\section{Introduction}

\begin{figure*}[t]
  \centering
  \includegraphics[width=\textwidth]{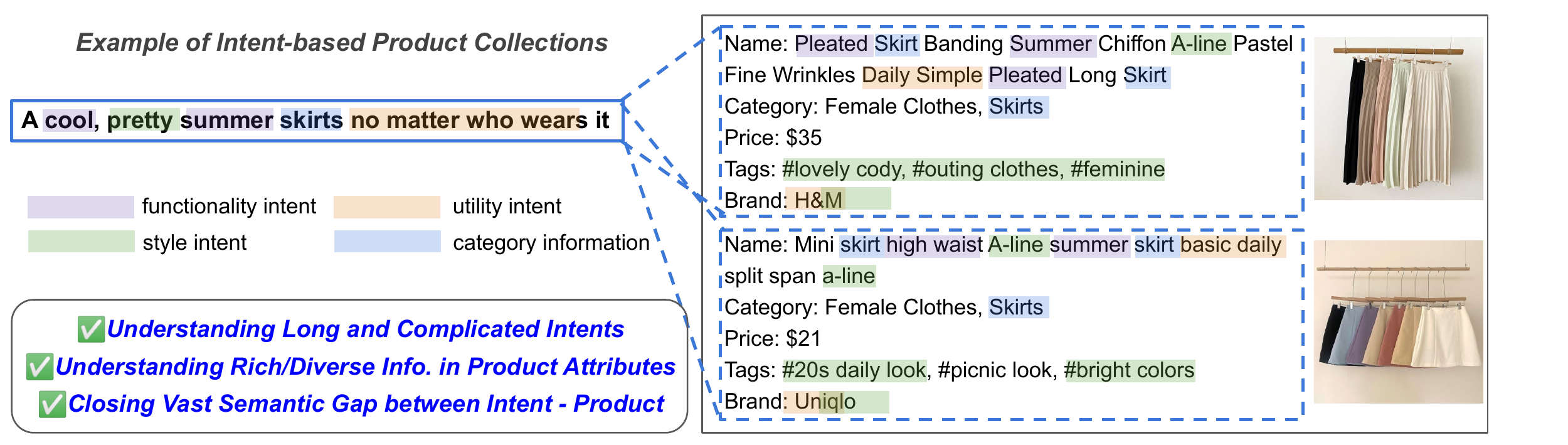}
  \caption{
  An example of constructing an intent-based product collection for e-commerce.
  }
  \label{comparison}
\end{figure*}

Current e-commerce platforms heavily rely on search engines or recommender systems to display their products to customers.
However, there have been limitations of such tools.
Customers do not precisely know which keywords they have to provide with search engines in advance.
Recommender systems only provide multiple products that are just top-k results of computation between user features and product features without explicit consideration of customer's intention~\cite{schafer2001commerce, wei2007survey, sivapalan2014recommender, shin2021one4all, 10.1145/3460231.3478844}.
Thus, unless the customer makes many iterative searches or browsing, their intention may not be satisfied.
As a solution, e-commerce platforms regularly display intent-based shopping product collections on their websites to provide convenient shopping experiences to users.
The assortment of products with a common theme can reduce the burden of iterative search or browsing by combining coherent but diverse products to fulfill customer's intentions at once.
Although there has been rapid progress in the recommender system domain, making product collection has still been a human task.
The quality of outcome is more prone to errors than that of a search engine or recommender system.
In other words, it is not yet an alternative solution to overcome the limitations of a current search engine or recommender system because both volumes and diversity of current product collections mostly retained in human scale.
Technically, there are three challenging problems to build a product collection automatically.
1) understanding of long and complicated intent sentences.
2) handling rich and diverse attributes of products combined in a collection.
3) closing a vast semantic gap between the customer's intention and the product's attributes.
Fig.~\ref{comparison} shows an example of constructing a collection. 
Typically, a single intent sentence contains multiple intents (e.g., functionality, utility, or style) as well as a product has rich and diverse textual attributes (e.g., `daily simple', `H\&M', `basic', or `Uniqlo') making the matching task complex.

In this paper, we build an intent-based product collection by using a pretrained language model.
BERT~\cite{devlin-etal-2019-bert} has powerful language understanding capability by pretrained knowledge and Self-Attention \cite{NIPS2017_transformers} to handle long sentences.
Specifically, we adopt Sentence-BERT (SBERT)~\cite{reimers-gurevych-2019-sentence} to fine-tune its parameters for matching textual attributes of products and corresponding intent sentences with triplet loss.
We set a concatenation of the title, section name, and date of a product collection as a query and up to eleven textual attributes of products in the collections as positive examples.
We use term-based search using the query for gathering hard negative examples.
Also, we augment positive pairs by modifying a query to contain category name.

Offline experimental results on five in-house datasets show that our negative sampling and data augmentation techniques improve recall and precision by a large margin.
Compare to the baseline search-based model, our model significantly improves the intent-based matching performance.
When performing online evaluations in our service, the product collections from the model increase the CTR, CVR, order-diversity by 16\%, 29\%, 60\%, respectively, comparing to the product collections made by human experts. 
We additionally present an ablation study by different sizes of the SBERT model.
Examples of intents and matched products using our model and an example of exposed product collection on our service are displayed in Appendix~\ref{matching-example} and~\ref{screenshot}, respectively.

\section{Related Work}
\label{related-works}

\subsection{Intent-based Product Collections}
Knowledge graphs are used to identifying products having similar attributes~\cite{a11090137, catherine2017explainable, autoknow}.
\cite{alicoco} built a huge knowledge graph whose node types are items, e-commerce concepts, primitive concepts, and taxonomy. And for a given specific concept, they collect products that are connected with the concept.
Motivated by existing bag-of-words approaches \cite{lsa, lda},
Angelov~\cite{angelov2020top2vec} proposed Top2Vec, a representation learning for topic modeling.
However, since their models treat an intent as one object, e.g. a node or a topic, their representation power of intents is limited.

\subsection{Metric Learning for Product Retrieval} Metric learning learns and measures the similarity of data instances, where a relevant representation of sentences can directly benefit performance \cite{metric-learning-survey-2019}.
Many representation learning techniques
are used for metric learning such as Bilinear model \cite{10.5555/2567709.2567742}, CNNs \cite{NIPS2014_b9d487a3, cnn-ir-2014}, RNNs \cite{rnn4ir}, and others \cite{dssm, drmm, hui-etal-2017-pacrr, 10.1145/3159652.3159689, hui2017re, mitra2016dual, 10.1145/3038912.3052579}.
In application to product search, Nigam et al.~\cite{semantic-product-search} introduced a shared embedding model for semantic product search. 
They focus on search queries, such as categories and names, that are typically short thus their approach also has a limitation for understanding and relating long, complicated intents to rich product attributes.

\subsection{Pretrained Language Models for Product Understanding}
BERT has been applied for general semantic relevances task~\cite{liu2021pre} and web search~\cite{sun2020ernie}
for its powerful language understanding capability compared with conventional lexical matching~\cite{bm25}.
For e-commerce tasks,
BERT has been applied to product specification understanding~\cite{roy-etal-2020-using} and user behavior prediction using session log-based product representations~\cite{tagliabue-etal-2021-bert}.
Reimers et al.~\cite{reimers-gurevych-2019-sentence} presented Sentence-BERT (SBERT) for similarity learning using siamese or triplet BERT architecture.

We choose SBERT for our task,
because SBERT could understand arbitrary intents and rich product attributes end to end without manual effort of building ontologies.
Even if metric learning for BERT has been used for product matching~\cite{tracz-etal-2020-bert},
applying SBERT to textual attributes of large, manually crafted collections to facilitate intent-based product retrieval is not intensively studied to the best of our knowledge.

\begin{figure*}[t]
  \centering
  \includegraphics[width=\textwidth]{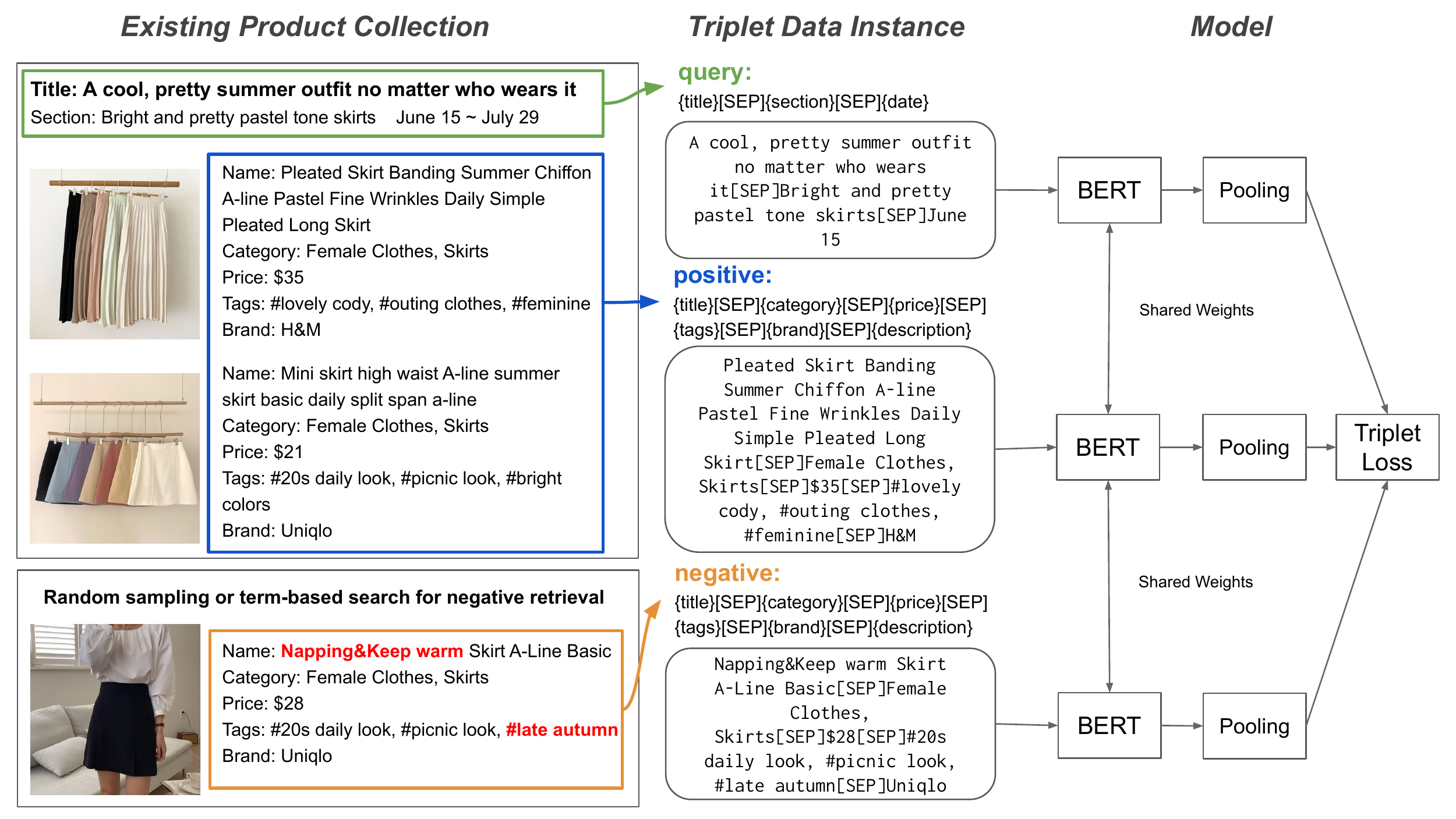}
  \caption{
  Illustration of data instances and model architecture used for intent-based product retrieval. %
  }
  \label{arch}
\end{figure*}

\section{Method}
\label{method}

\subsection{Task Description}
\label{task}

When shopping products are visible at a glance, it is easy to compare and choose the product for specific themes or purposes. For this reason, many e-commerce platforms, such as Amazon, Shopify, and Zalando, collect the products under certain themes or purposes and expose them to their main page. Because of its importance, this work is usually done by human experts. Human experts figure out various shopping intents of users and make product collections that users can satisfy.
However, they spend too much time and have a lot of difficulties choosing the proper products across diverse themes and purposes of shopping. This is the motivation of our work. We have a sentence, which includes users' shopping intent, as an input and then output the collection of products with a common intent.
Fig.~\ref{comparison} emphasizes the advantage of intent-based product retrieval.
Formally speaking,
we have a set of products $P_{full}$ and a shopping intent $I$,
then our task is to match a subset of products $P_{sub}$ from $P_{full}$ with $I$.
\begin{equation}
\label{eq0}
P_{sub} = Matching(P_{full}, I)
\end{equation}

\subsection{Metric Learning with Product Collections Data}

We adopted SBERT for implementing $Matching$ function in Eq.~(\ref{eq0}).
We reconstruct existing product collection data created by human operators for metric learning with triplet loss~\cite{triplet-loss}.

\subsubsection{Training Objectives and Loss Functions}
The training objective is to create representation for a given intent query and positive products to have high similarity while keeping low similarity between the query and negative product.
On the right side of Fig. \ref{arch} shows neural network architecture corresponds to anchor, positive, and negative examples.

Existing product collections consist of triplet anchor (query), and positive products as in the left side of Fig.~\ref{arch}.
For negative products, we sampled randomly or retrieved from the term-based search results. %

For a given query $Q$, a positive product $P$ and a negative product $N$,
let $e_Q$, $e_P$, and $e_N$ be an embedding vector of $Q$, $P$, and $N$, respectively, after applying BERT and pooling. 
For triplet loss for metric learning, %
we use Euclidean norm and triplet margin to 1 followed by~\cite{reimers-gurevych-2019-sentence}. The formula is described as follows: %
$$
Loss(Q, P, N) = ReLU(\| e_Q - e_P\| - \| e_Q - e_N\| + Margin)
$$

\subsection{Constructing BERT Input of Intents and Products}
A single product collection is decomposed to 50$\sim$100 triplet positive pairs for training.

\subsubsection{Intent Query}
The intent query is composed of a title, section name (name of products group in the product collection), and start date of the product collection.
For example,
title of product collection in Fig.~\ref{arch} `A cool, pretty summer outfit no matter who wears it' is an intent that describes the functionality, utility, or style.
Products inside that product collection have high conformance for that intent.
The section name `Bright and pretty pastel tone skirts' of the product collection involves both an intent and product category (i.e. skirts).
If category information exists in the section name, product collection needs to be composed of products from that category.
The product collection's start date of exposure on the e-commerce platform is also used as part of a query.
This information indicates seasonality.
For example, `June 15' is added to the query for the model to possibly learn representations for relevant products of early summer (e.g. skirts with light color). %
The elements of the query are concatenated with \texttt{[SEP]} token.
\subsubsection{Positive and Negative Products}
The product is represented as a sentence containing up to eleven textual product attributes such as title, categories, price, tags, or brand concatenated with \texttt{[SEP]} token.
The positive product is derived from the existing product collection with corresponding intent query.
The negative product is randomly sampled or retrieved with BM25 term-based search~\cite{bm25} by the query text.
The purpose of the BM25 term-based search is to gather hard negatives effectively~\cite{xiong2020approximate}.

\subsection{Joint Learning with Category-Wise Data Augmentation}
\label{joint-training}

\subsubsection{Overlapped Intents across Categories}
Products in intent-based collections usually share the same or similar product category.
For instance, Fig.~\ref{comparison} shows the intent of `A cool, pretty summer skirts no matter who wears it'.
This intent contains category information `skirts'.
Since we are training with many existing product collections datasets,
there will be product collections entitled
`A cool, pretty summer pants no matter who wears it', or `A cool, pretty summer shirts no matter who wears it'.
Therefore, the model has a chance to match the wrong products (pants and shirts) along with skirts since their pure intent sentences are similar.

In addition, experts also include different category products at the mid or later part of product collections for complementary products displays (e.g. displaying skirts and pants together). 
However, our training data do not consider the significance of the relative order of products in a collection.
Thus, queries may have wrong category information to the portion of products in a collection (e.g. skirts query mapped to pants products).
A model trained with this data can potentially result in categorically wrong product retrieval.

\subsubsection{Category-Wise Data Augmentation}
\begin{figure}
  \centering
  \includegraphics[width=\linewidth]{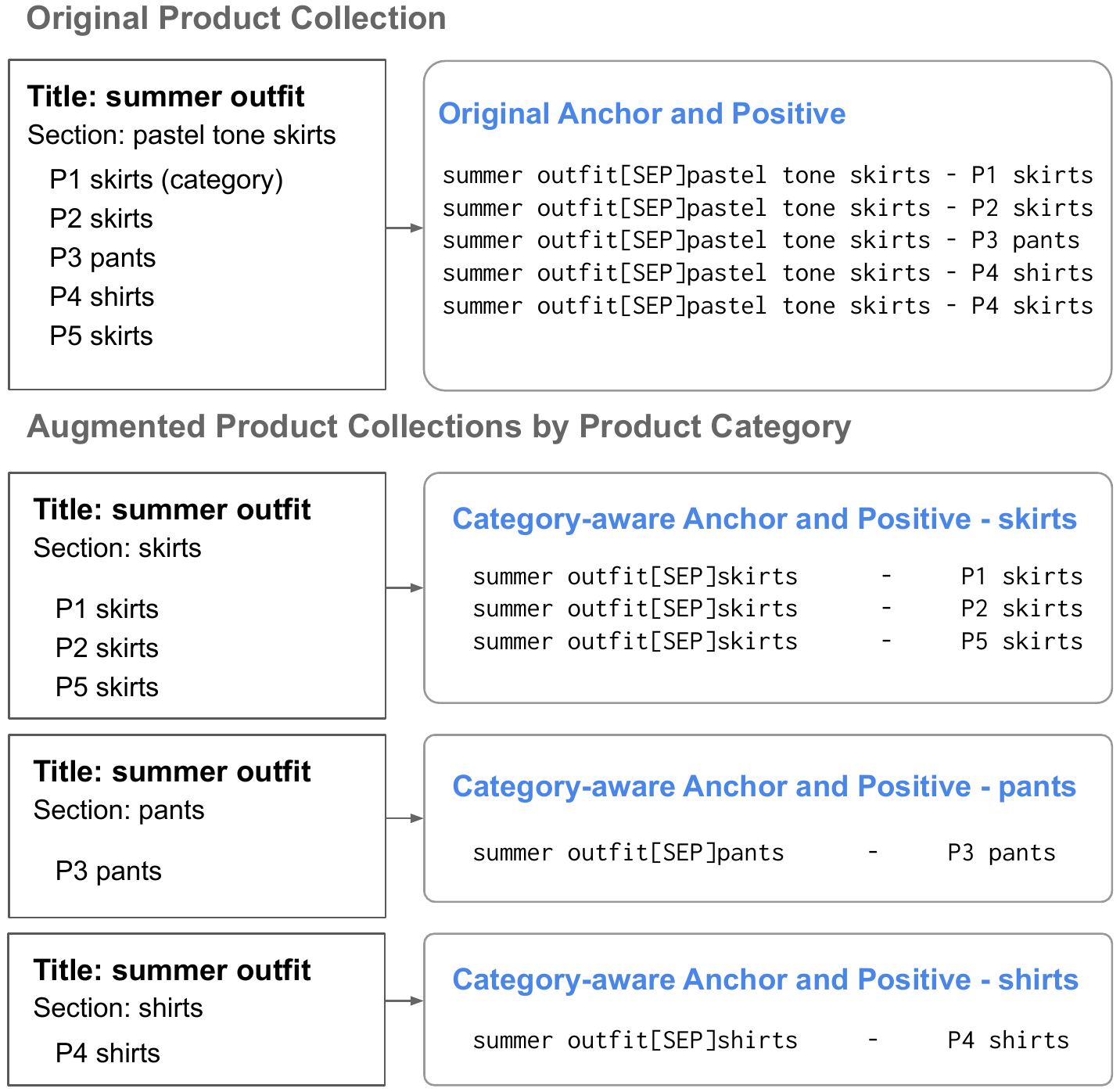}
  \caption{Example of category-wise data augmentation.
  }
  \label{augmentation}
\end{figure}

To alleviate the categorical correctness issue, we perform simple data augmentation as visualized in Fig.~\ref{augmentation}.
1) Randomly select a specific ratio of product collections from all product collections in the training data.
2) Transform a single product collection that has multiple product categories into multiple product collections with a single product category. %
3) Replace a section name in the query of transformed product collections by common category name of products.
This simple augmentation 
forces model to more aware category information in the query,
and it significantly improves categorical precision, compared to the tradeoff made in the recall.
The detailed experimental result will be discussed in Section~\ref{experiments}.

\subsection{Model and Learning}

\subsubsection{Models}

Pretrained language model BERT is used for making product representations for various e-commerce tasks \cite{tagliabue-etal-2021-bert, roy-etal-2020-using, tracz-etal-2020-bert},
and it plays a baseline model for the modern intent understanding of natural languages in the dialog systems domain \cite{chen2019bert, wu-etal-2020-tod}.
Furthermore, BERT is plausible for joint learning \cite{chen2019bert}, which is helpful for the training model with augmented data aiming for both intent-based retrieval as well as categorically correct retrieval.
For these reasons, we adopt BERT as a base module for understanding products and intents.

For full model, we adopt SBERT \cite{reimers-gurevych-2019-sentence},
a siamese and triplet network version of the BERT, to project intent and product representation in the same vector space.
We used custom pretrained BERT weight (110M, 12L, 768 output dim.) for SBERT. We used the Wordpiece tokenizer \cite{wordpiece} for both pretraining and fine-tuning the BERT model to understand intent and product attributes with semantically relevant subwords.

\subsubsection{Architectural Analysis for Intent-based Product Retrieval}
BERT recognizes \texttt{[SEP]} token to separate different attribute types of a product.
After the model ingests the triplet dataset, Self-Attention \cite{NIPS2017_transformers} captures a latent representation of query ($e_Q$) and text information of product ($e_P$).
Self-Attention helps to make relevant representations for long and complicated intent queries.
It also helps to make relevant representations for arbitrary length product attributes by handling long-range dependencies.
Layer normalization \cite{ba2016layer} in Self-Attention alleviates potential length difference between query and product where the concatenated product attributes are usually a few times longer than an intent query.
We used the default \texttt{MEAN}-strategy pooling in SBERT to create output representation.
For production, we perform a cosine similarity search on top of $e_Q$ and precomputed $e_P$.

\begin{table}[t]
    \centering
    \caption{
    Statistics of product collections for training data.
    }
    ~\\[3.5pt]
    \setlength\tabcolsep{4pt}
    \label{tab:collection-stats}
    \begin{tabular}{l|c}
    \toprule
        Data & Counts\\
        \midrule
       Collections & 88,371\\
       \midrule
       Sections in Collections & 139,794\\
       \midrule
       Products in Collections & 8,790,251\\
       \midrule
       Avg. Product per Collection & 99.46\\
       Avg. Product per Section & 62.88\\
       \midrule
       Total Product Categories & 1,154\\
       \bottomrule
    \end{tabular}
\end{table}

\begin{figure}[t]
\centering

\includegraphics[width=\linewidth]{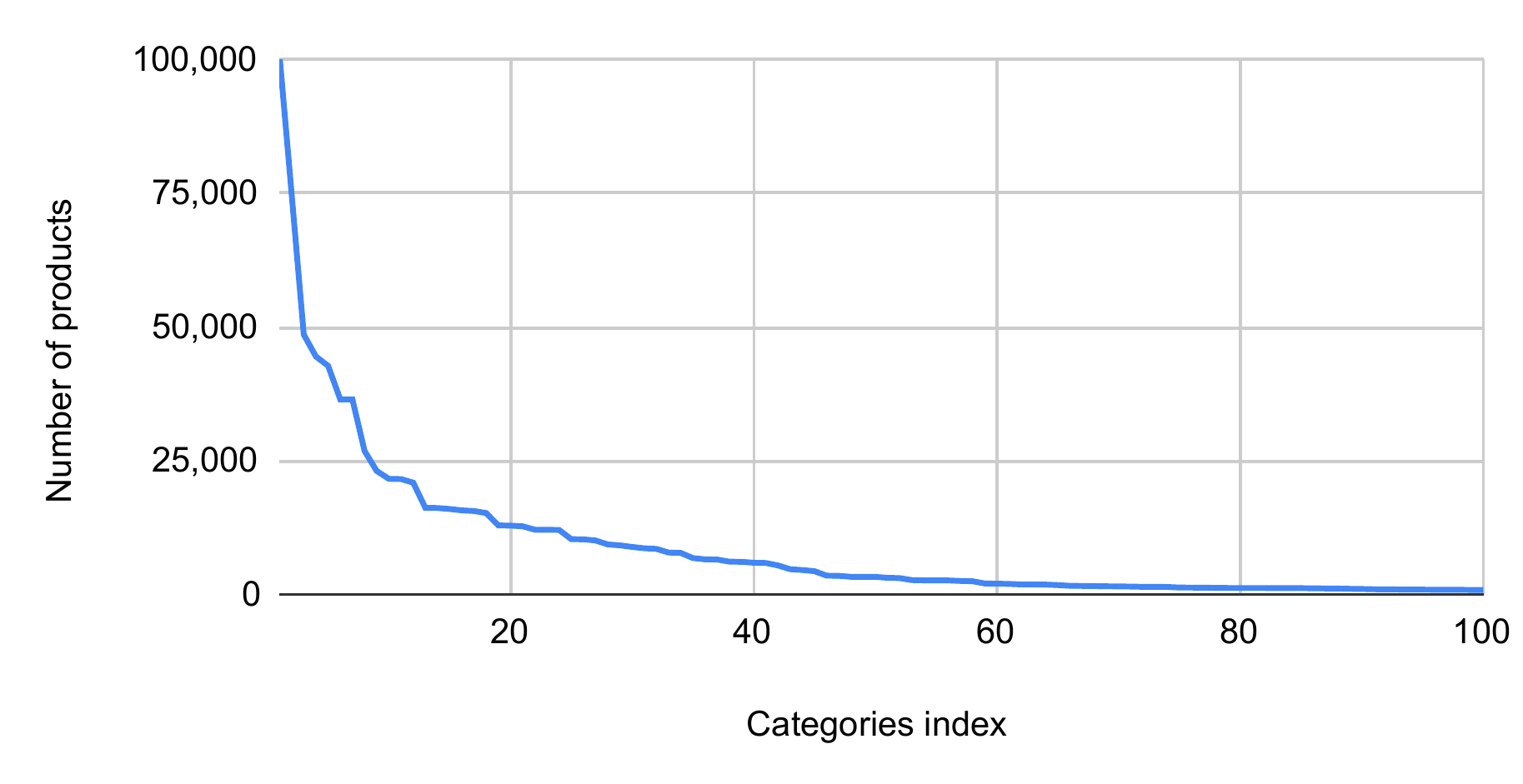}
  \caption{The numbers of positive products belong to the top 100 most frequent categories.}
  \label{category-dist}
\end{figure}

\begin{table*}[t]

\begin{minipage}[c]{0.7\linewidth}
\centering
    \caption{
    Dataset configuration and statistics of models.
    }
    \setlength\tabcolsep{4pt}
    \label{tab:training-datastat}
    \begin{tabular}{c|cc|c|cccc}
    \toprule
       \multirow{2}{*}{Model} & \multicolumn{2}{c|}{Negative Sampling}  & \multirow{2}{*}{\shortstack{Cat.-Wise\\Data Aug.}} & \multirow{2}{*}{\shortstack{Products for\\Positive Pairs}} & \multirow{2}{*}{\shortstack{Products for\\Negative Pairs}} & \multirow{2}{*}{\shortstack{Total Triplet\\Data Instances}}\\
       & Product Cat. & BM25 & \\
       \midrule
        $PR_{easy0}$ & 25 & 0 & 0\% & \multirow{5}{*}{968,202} & 968,202 & 73,752,535 \\
        $PR_{hard0}$ & 10 & 15 & 0\% & & 6,477,564 & 62,307,766 \\
        $PR_{hard15}$ & 10 & 15 & 15\% & & 6,363,214 & 67,266,919 \\
        $PR_{hard40}$ & 10 & 15 & 40\% & & 6,060,951 & 78,870,990 \\
        $PR_{hard55}$ & 10 & 15 & 55\% & & 5,976,943 & 87,502,182 \\
       \bottomrule
    \end{tabular}
\end{minipage}
\begin{minipage}[c]{0.29\linewidth}
    \centering
    \caption{
    The number of products belonging to each category in the evaluation data.
    }
    \setlength\tabcolsep{4pt}
    \label{tab:eval-data}
    \begin{tabular}{c|c}
    \toprule
        Category & Product Counts\\
        \midrule
        Underw. & 148,118 \\
        Bags & 212,531 \\
        Accs. & 476,229 \\
        Goods & 225,799\\
        Shoes & 326,109\\
       \bottomrule
    \end{tabular}
    
\end{minipage}

\end{table*}

\section{Experiments}
\label{experiments}

\subsection{Training and Evaluation}

\subsubsection{Training Data}
We extracted product collections for the fashion category published from in 2.5 years period.
The statistics of extracted product collections is stated in Table~\ref{tab:collection-stats}.
Due to the operation policy of our platforms,
information of some old products are not available.
Thus, only currently accessible products (`Products for Positive Pairs' in Table~\ref{tab:training-datastat}) are used for training.
The category distribution of products is depicted in Fig.~\ref{category-dist}.

With the same query and positive products from product collection data,
we apply different negative sampling settings and category-wise data augmentation ratios to make five models.
Table~\ref{tab:training-datastat} depicts models with dataset configuration and statistics.
The model name represents the negative sampling strategy and the ratio of category-wise data augmentation.
For a single positive product, we use two types of negative samples.
First, random sampling from the same categories, and second, sampling by BM25 Search using the query from product collections. We denote $_{hard}$ in the model name when involves BM25-based negative samples and $_{easy}$ for random sampling.
We augment collections category-wise at a specific ratio from all product collections.
We denote the augmentation percentage to the model name (e.g. $_{40}$).
The model names will connect evaluation results to the corresponding dataset configuration.

\begin{table*}[t]

\begin{minipage}[c]{\linewidth}
 \centering
    \caption{
    Recall and precision evaluation results using BM25 Search.
    }
    \setlength\tabcolsep{2.6pt}
    \label{tab:baseline_metrics}
    \begin{tabular}{cccccc|cccccc}
    \toprule
      \multicolumn{6}{c|}{Recall@100 - BM25 Search}  & \multicolumn{6}{c}{Precision@100 - BM25 Search} \\
      Underw. & Bags & Accs. & Goods & Shoes & Avg. & Underw. & Bags & Accs. & Goods & Shoes & Avg.\\
      \midrule
        0.0378&	0.0341&	0.0211&	0.0403&	0.0273&	0.03212&	0.5363&	0.6172&	0.7797&	0.1662&	0.7261&	0.5651\\
      \bottomrule
    \end{tabular}
    ~\\[10pt]
    \caption{
    Recall evaluation results of models at 400,000 training steps.
     The best score is \textbf{bold} and second best score is \underline{underlined} among models with hard negatives and category-wise data augmentation ($PR_{hard15}$, $PR_{hard40}$, $PR_{hard55}$).
    }
    \vspace{6px}
    \setlength\tabcolsep{2.6pt}
    \label{tab:recall}
    \begin{tabular}{c|cccccc|cccccc}
    \toprule
      \multirow{2}{*}{Model} & \multicolumn{6}{c|}{Recall@100 - SBERT 12L}  & \multicolumn{6}{c}{Recall@100 - SBERT 6L} \\
      & Underw. & Bags & Accs. & Goods & Shoes & Avg. & Underw. & Bags & Accs. & Goods & Shoes & Avg.\\
      \midrule
        $PR_{easy0}$ & 0.5141 & 0.4840 & 0.5122 & 0.5032 & 0.4156 & 0.4858 & 0.4733 & 0.4163 & 0.4920 & 0.4983 & 0.4044 & 0.4569\\
        $PR_{hard0}$ & 0.5830 & 0.4493 & 0.5776 & 0.5983 & 0.4433 & 0.5303 & 0.5505 & 0.4106 & 0.5440 & 0.5698 & 0.3786 & 0.4907\\
      \midrule
        $PR_{hard15}$ & \textbf{0.5847} & \textbf{0.4374} & \textbf{0.5724} & \textbf{0.5874} & \textbf{0.4331} & \textbf{0.5230} & \underline{0.5859} & \textbf{0.4026} & \textbf{0.5535} & \textbf{0.5781} & \underline{0.4088} & \textbf{0.5058}\\
        $PR_{hard40}$ & \underline{0.5816} & \underline{0.3601} & \underline{0.5564} & \underline{0.5583} & \underline{0.4192} & \underline{0.4951} & 0.5851 & \underline{0.3773} & \underline{0.5340} & \underline{0.5359} & \textbf{0.4151} & \underline{0.4895}\\
        $PR_{hard55}$ & 0.5659 & 0.3459 & 0.5215 & 0.5286 & 0.3930 & 0.4710 & \textbf{0.6169} & 0.3154 & 0.5227 & 0.5262 & 0.3914 & 0.4745\\
      \bottomrule
    \end{tabular}
    ~\\[10pt]
    \caption{
    Precision evaluation results of models at 400,000 training steps.
     The best score is \textbf{bold} and second best score is \underline{underlined}.
    }
    \vspace{0px}
    \setlength\tabcolsep{2.6pt}
    \label{tab:precision}
    \begin{tabular}{c|cccccc|cccccc}
    \toprule
      \multirow{2}{*}{Model} & \multicolumn{6}{c|}{Precision@100 - SBERT 12L}  & \multicolumn{6}{c}{Precision@100 - SBERT 6L} \\
      & Underw. & Bags & Accs. & Goods & Shoes & Avg. & Underw. & Bags & Accs. & Goods & Shoes & Avg.\\
      \midrule
      $PR_{hard15}$ & 0.2877 & 0.6101 & 0.5334 & \textbf{0.1810} & 0.4928 & 0.4210 & 0.2861 & \textbf{0.6589} & 0.5382 & \textbf{0.1821} & 0.5506 & 0.4432\\
      $PR_{hard40}$ & \underline{0.4711} & \textbf{0.6717} & \underline{0.6288} & 0.1547 & \underline{0.6832} & \underline{0.5219} & \underline{0.4529} & \underline{0.6579} & \underline{0.6695} & 0.1536 & \underline{0.6698} & \underline{0.5207}\\
      $PR_{hard55}$ & \textbf{0.5305} & \underline{0.6447} & \textbf{0.6929} & \underline{0.1715} & \textbf{0.7625} & \textbf{0.5604} & \textbf{0.5073} & 0.6572 & \textbf{0.7060} & \underline{0.1715} & \textbf{0.6831} & \textbf{0.5450}\\
      \bottomrule
    \end{tabular}
\end{minipage}%
\end{table*}

\subsection{Metric Analysis}

\subsubsection{Evaluation Data and Protocol}
The evaluation procedure for product retrieval systems is usually divided into two-phase: 1) indexing embeddings from evaluation products (ideally, same as online products) 2) making query embedding and retrieve products from evaluation indices.
For queries, we used the same product collection data used for training to make queries for evaluation.
For products, we used extract products with specific customer review counts (in this case same or larger than 1) from online products.
In our e-commerce platform,
the number of online products in the major categories reaches tens of millions, it is not feasible to perform the evaluation on top of entire online products.
Hence we used this filtering condition to make a small product volume for evaluation while maintaining extensive coverage of products across product collections.
We only used product collections contains five or more evaluation products for recall evaluation.
The statistics of evaluation data are depicted in Table~\ref{tab:eval-data}.

\subsubsection{Evaluation Metrics}

In a single product collection dataset,
we evaluate how well trained SBERT model retrieves and restores original products by a given query (i.e. recall to measure intent conformance of retrieval result).
If data is category-wise augmented,
we perform an additional evaluation for how well trained SBERT model retrieves products with the same category mentioned in a given query (i.e. precision to measure categorical correctness of retrieved result).
Formally speaking,
For a set $P_{gt}$ of evaluation products in the specific product collections,
a recall is defined as follows:
$$
recall = \cfrac
{\text{the number of retrieved products} \in P_{gt}}
{\text{the number of } P_{gt}}
$$
For a set $C_{gt}$ of products with ground truth category in query, a precision is defined as follows:
$$
precision = \cfrac
{\text{the number of retrieved products} \in C_{gt}}
{\text{the total number of retrieved products}}
$$

In measuring recall and precision,
the number of retrievals performed for evaluations is different by the ratio of category-wise augmentation of each model training (0$\sim$55\% in this paper).
However, we assume the relative difficulties of retrieval tasks for all model training are mostly the same because only queries used for training will be used for evaluation.

We train the model up to 400,000 steps and the evaluation interval is 20,000 steps.
The training batch size is 55.
We used P40 GPU for most model training.
We used Faiss \cite{faiss} for similarity search for evaluation.

\subsubsection{BM25 Search Baseline}
Table~\ref{tab:baseline_metrics} represents the recall and precision performance of the baseline BM25 Search.
For query in BM25 Search model evaluation, we additionally filter uninformative words in the title (e.g. `collection') since some of our product collections are entitled with the word `collection'.
We also exclude the date information for the query, since many collections share the same date and potentially lead to noise in retrieval results.
We tokenize query and product attributes by space.
All other details on evaluation are the same as SBERT models.

\subsubsection{Metric Overview}
Table~\ref{tab:recall} and Table \ref{tab:precision}
show recall and precision performance of SBERT models in various product categories.
The metrics of models at each training interval are depicted in Appendix \ref{recall-precision-graph}.
Since models require optimal recall and precision performance for our task, we marked the best score and second best score only among $PR_{hard15}$, $PR_{hard40}$, $PR_{hard55}$ models (i.e. models with hard negative samples and category-wise data augmentation).
We conduct experiments in SBERT 12L and 6L model, here we focused on the performance of the 12L model, a comparison between the 12L and 6L model will be discussed in Section~\ref{ablation-study}.

\subsubsection{Recall}
Table \ref{tab:recall} depicts recall performances.
$PR_{hard0}$ is the best by adopting hard negative sampling without category-wise data augmentation, nearly 9.2\% recall improvements %
for between $PR_{easy0}$ and $PR_{hard0}$.
This means hard negative sampling using the query of product collection is effective for triplet metric learning.
We expect
these hard negative examples not only considering product category but also considering similar but undesirable product attributes such as specific material or color differences. 
This helps the model to learn rich and unique product representations of the given intent query.

$PR_{hard15}$ and $PR_{hard40}$ are the second and third best performances, respectively. This means category-wise data augmentation degrades recall performance. 
Since category information is more common compared to the intent sentence in the dataset,
adopting category information directly to query (intent) embedding forces model to create more general query representations.
Recall depends on unique representations of queries, thus
general query representations degrade recall performances.

The left side of Table~\ref{tab:baseline_metrics} depicts recall performance of baseline BM25 Search.
Since BM25 Search only considers word-level matching, it fails to map long and complicated queries to rich product attributes, and introduces very low recall performance.

\subsubsection{Precision}
Table \ref{tab:precision} shows precision performance.
We observe the percentage of category-wise augmented data correlated to precision performance $PR_{hard55}$ $>$ $PR_{hard40}$ $>$ $PR_{hard15}$.
The best model matches product from the correct category by 56\% of product collections, nearly 33.2\% precision improvements between $PR_{hard15}$ and $PR_{hard55}$, which means joint learning with category-wise positive pair augmentations helps the model to consider category-based precision along with intent matching.

The right side of Table~\ref{tab:baseline_metrics} depicts precision performance of baseline BM25 Search.
In many cases, category information is explicitly stated in product attributes, thus word-level matching can be enough to capture. The performance of BM25 Search is similar to $PR_{hard55}$, in other words, pretrained language models successfully satisfy categorical conditions while maintaining significantly improved recall performance.

\subsubsection{Optimal Fine-Tuned Model for Production}
We choose $PR_{hard40}$ for our services because the precision score and the recall score are balanced.
Even if the recall of $PR_{hard40}$ is 5.3\% less than the one of $PR_{hard15}$,
the precision of $PR_{hard40}$ is 24\% higher than the one of $PR_{hard15}$.
Compare to BM25 Search, the precision of $PR_{hard40}$ is slightly lower. However, the recall of it is outperformed significantly.
Note that, in a service environment, we apply post-processing for the more improved precision performance for SBERT models.

\section{Ablation Study}
\label{ablation-study}

Along with 12L (110M) model, we conduct an ablation study with the 6L (68M) model to analyze how the number of transformer layers and model capacity affects recall and precision performance.

We choose the 12L model for our services by the reliable recall and precision performance in both intent-based and category-based product retrieval.
For example, 
the 6L model is better on the precision performance of $PR_{hard15}$, but the 12L model is better on recall performance on a similar margin.
This is meaningful since the increase in the recall is generally harder than the increase in the precision in our task in terms of retrieval search space.
In addition, the 6L model shows slightly better recall performance to the 12L model on $PR_{hard55}$, but precision performance under the 12L model, which implies that adding more transformer layers is helpful for precision performance.
Furthermore, for $PR_{hard40}$, both recall and precision of the 12L model is higher than those of the 6L model.

\section{Online Performance}
\label{online-evaluation}

\begin{figure}[t]
  \centering
  \begin{minipage}[c]{\linewidth}%
  \includegraphics[width=\linewidth]{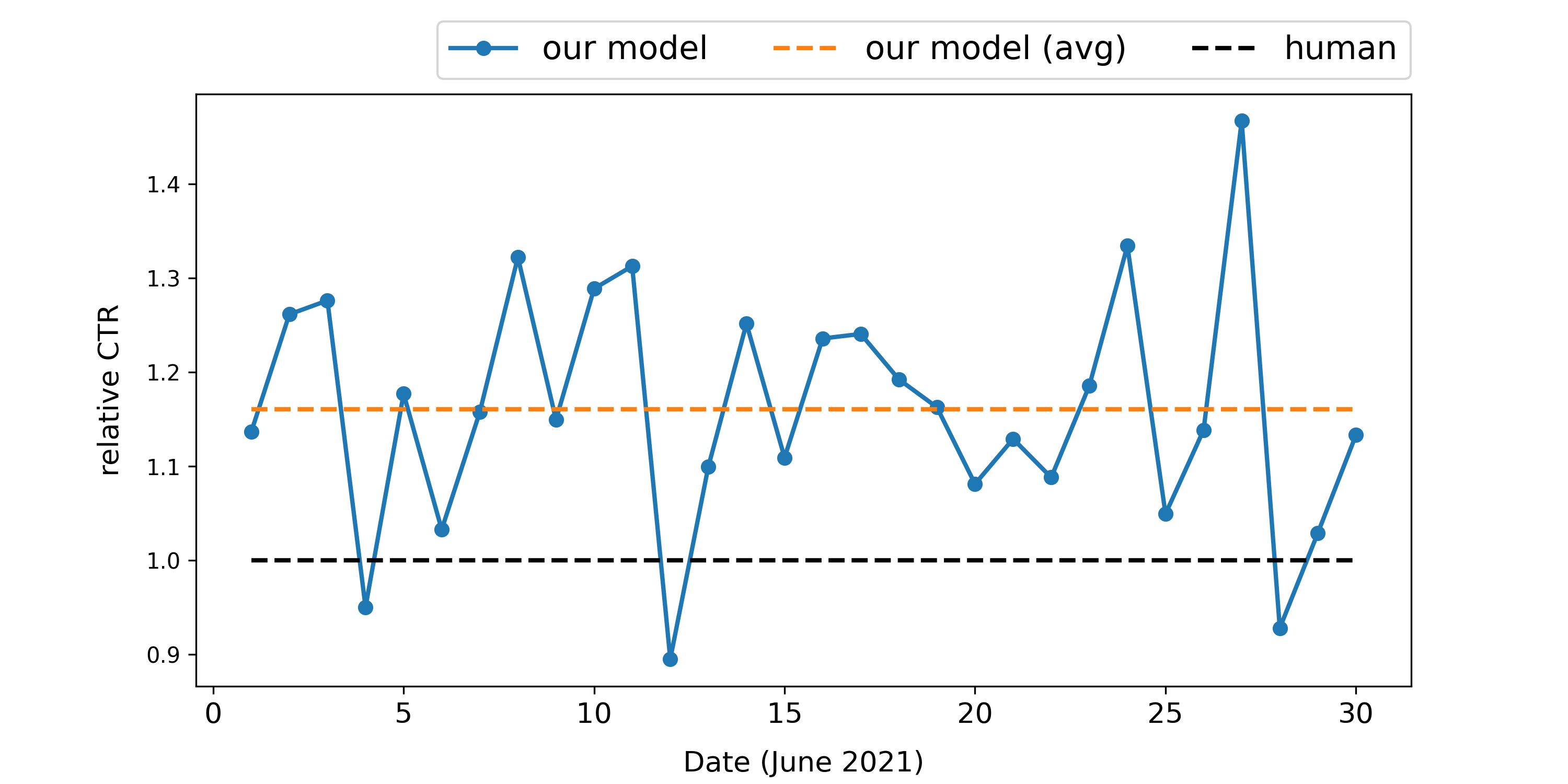}
  \end{minipage}\hfill
  \begin{minipage}[c]{\linewidth}%
  \includegraphics[width=\linewidth]{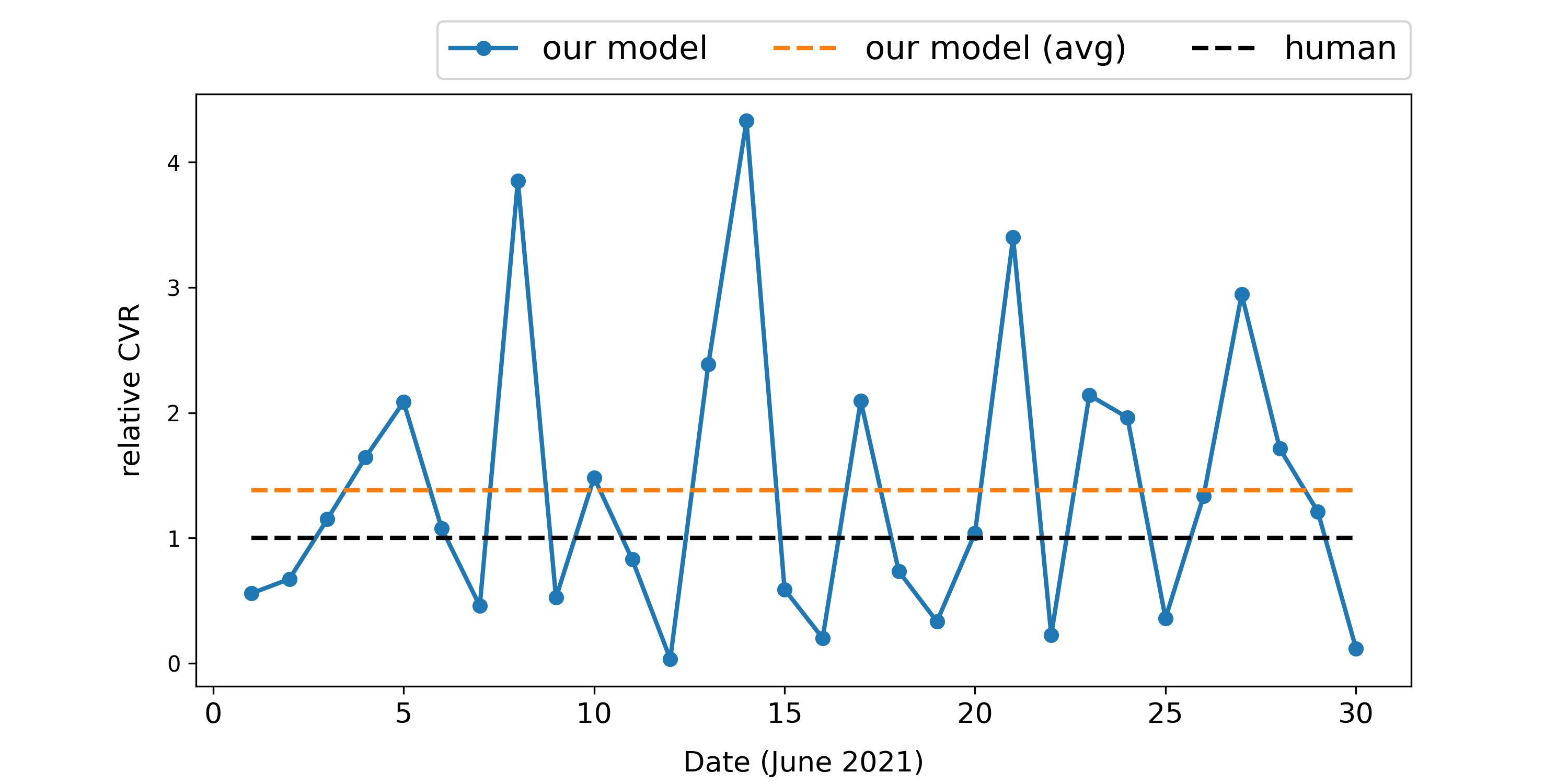}
  \end{minipage}\hfill
  \begin{minipage}[c]{\linewidth}%
  \includegraphics[width=\linewidth]{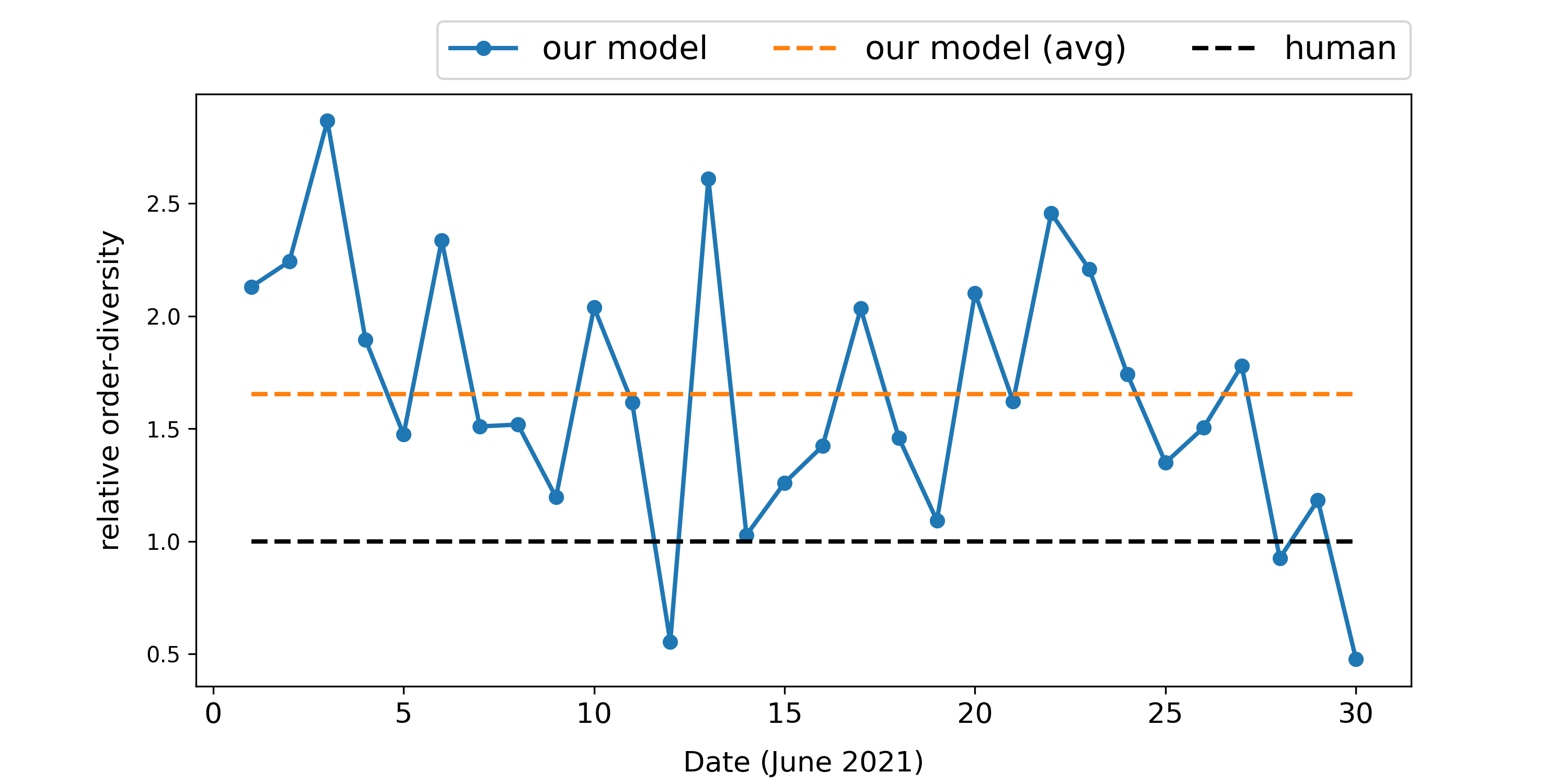}
  \end{minipage}
  \caption{Daily relative performance of our model. The x-axis means the date in June 2021. The orange dash line means the average relative CTR, CVR, and order-diversity. Overall, comparing to the product collections made by human operators, the product collections made by our model increase the CTR, CVR, order-diversity by 16\%, 29\%, 60\%, respectively.}
  \label{fig:online}
\end{figure}

In our service, %
we collect products under certain themes and show them to customers.
Currently, making the collections of products is done by dozens of professional human operators.
They first think of the theme from their own knowledge,
and then repeatedly use lexical search to gather related products.
We apply our model to automatically retrieve relevant products from a given theme query.

In this section, we compare the online performances of manually created product collections and product collections crafted using our model.
For each product collection, we compute CTR (Click Through Rate), CVR (Conversion Rate), and the order-diversity.
And then we compute the relative score as follows:
$$
relative\ score = \frac{\text{ avg. score of collections using our model }}{\text{ avg. score of collections made by experts }} 
$$
Here, CTR is the number of product clicks over the number of product collection views
and CVR is the number of product purchases over the number of product collection views.
The order-diversity, which can measure the diversity of purchased items, is the number of purchased products 
over the number of products in the product collection.
Thus, we can say that if the order-diversity of a product collection is large, 
then it contains a lot of quality products that are good enough to be purchased 
among web-scale products.

Comparing to the product collections made by humans,
the product collections made by our model increase the CTR, CVR, order-diversity by 16\%, 29\%, 60\%, respectively, in June 2021. 
The daily relative performance is illustrated in Fig.~\ref{fig:online}.
Using pretrained language models, it is possible to learn the product features
and collect related products with a common intent well.
For these reasons, our model consistently outperforms human operators, 
who are prone to make the product collection with their individual preferences and interests.

Note that when we make a product collection,
we first retrieve a bunch of products and reorder them.
Human experts usually focus on popularity or the number of reviews.
In our model, we run a simple linear regression model with similar features for reordering.

\section{Conclusion and Future Work}
\label{conclusion}

In this paper,
we create an intent-based product collection using SBERT that well performs in offline and online evaluation results to enable a better shopping experience for e-commerce platform users.
We adopt SBERT for triplet metric learning for creating intent-based product collections by the deep understanding of intent and related product attributes.
We enhance the base model by hard negative sampling for improved intent-based recall performance and category-wise positive pair augmentation for improved category-based precision performance.
We analyze effective training data construction options for fine-tuning, and ablation study to identify optimal pretrained model choices for production.
Furthermore, we measure online performance on CTR, CVR, order-diversity, and conclude that our model can create better intent-based product collections compare to human operators.
Currently, we are training and inferencing our model for most e-commerce categories including digital, living, or food.
In the future, we will conduct more detailed experiments on other categories.
We will also explore approaches for more effective model training and more personalized collections.

\IEEEpeerreviewmaketitle

\section*{Acknowledgment}
The authors would like to thank Jooho Lee, Hyun Ah Kim, Homin Ahn, Jin Ju Choi, Jinuk Kim, Minyoung Jeong, Nako Sung, Seungjae Jung, Jung-Woo Ha, and all of the project members in NAVER for devoted support and discussion.
Finally, the authors thank anonymous reviewers of DMS2021 and FashionXRecsys21 for their valuable comments.

\bibliographystyle{IEEEtran}
\bibliography{base}

\appendices %

\clearpage

\begin{table}
\section{Recall and precision metrics}
\label{recall-precision-graph}

\begin{center}
\includegraphics[width=.824\linewidth]{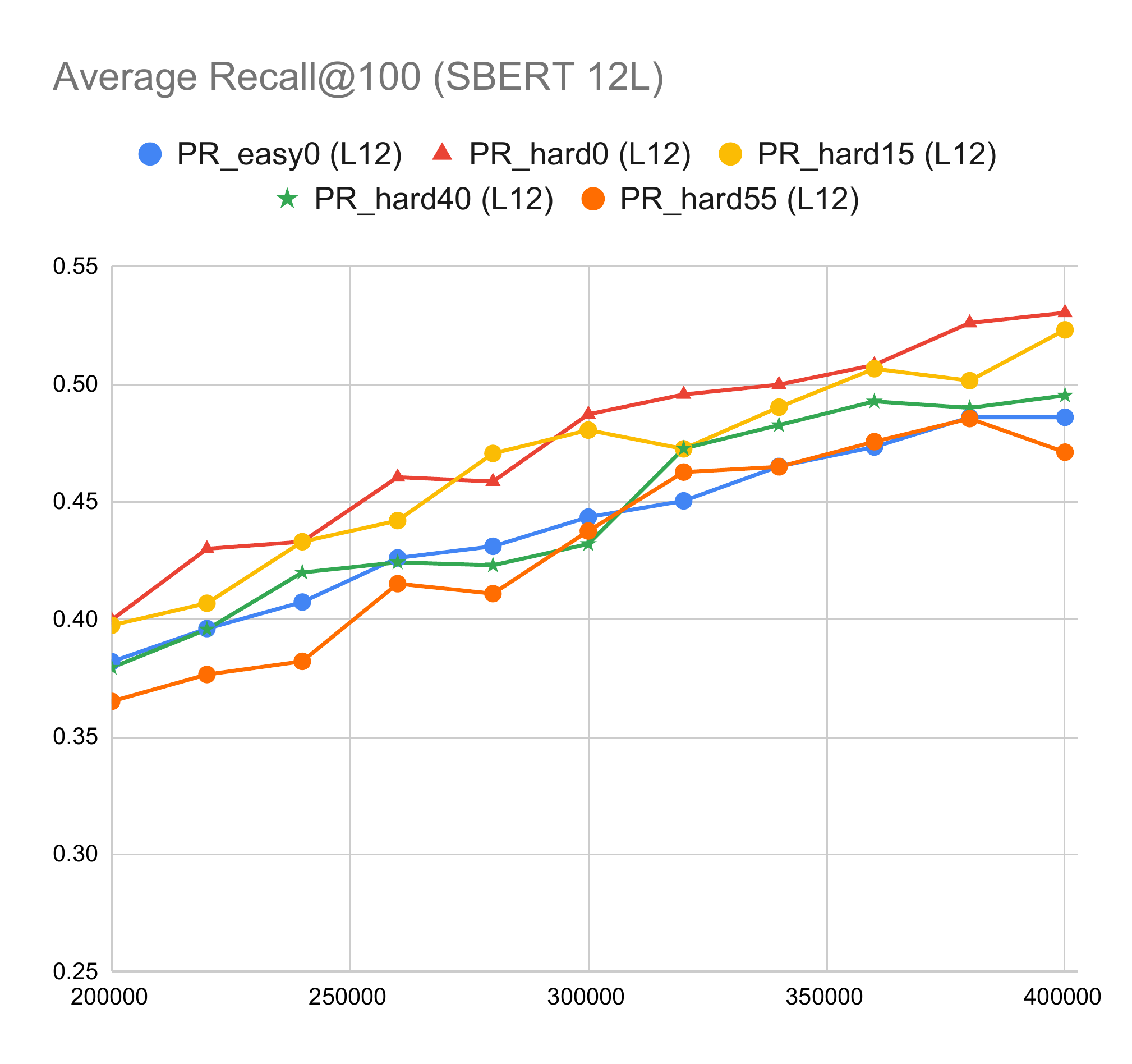}
\includegraphics[width=.824\linewidth]{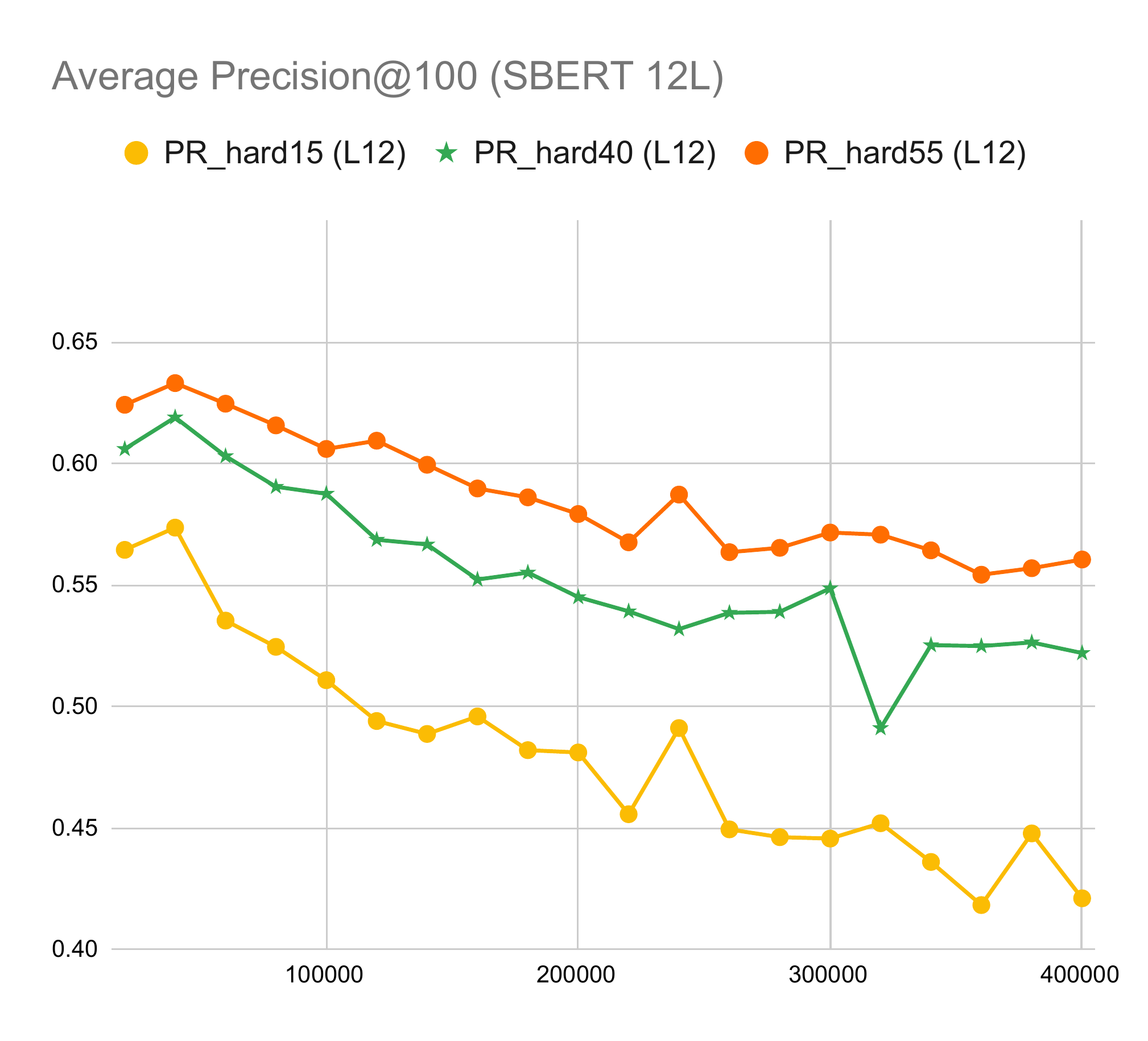}
\end{center}
\end{table}

\section{Examples of intents and matched products}
\label{matching-example}
~\\[-20pt]
\begin{center}
\centering
\includegraphics[width=\linewidth]{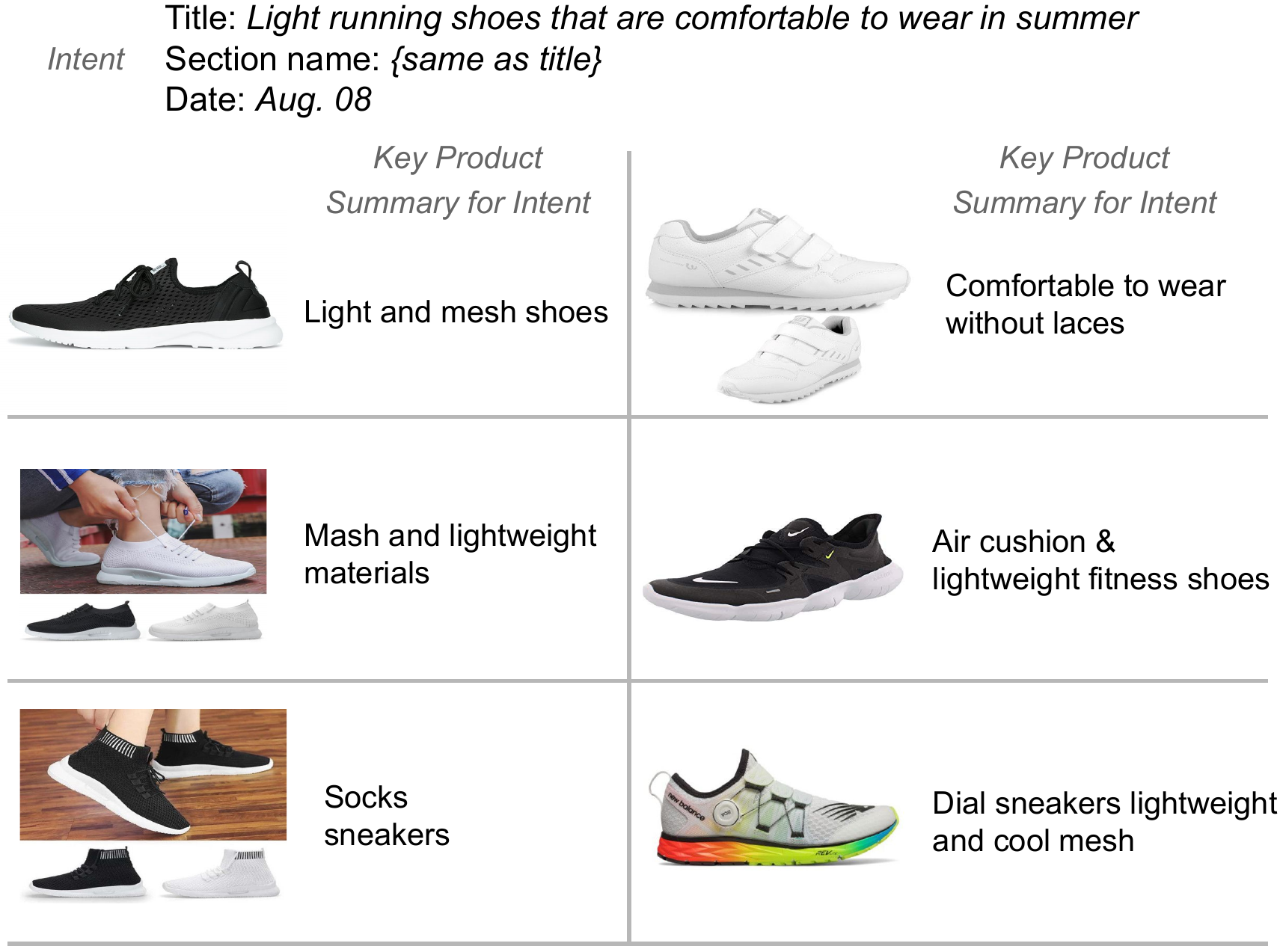}
\includegraphics[width=\linewidth]{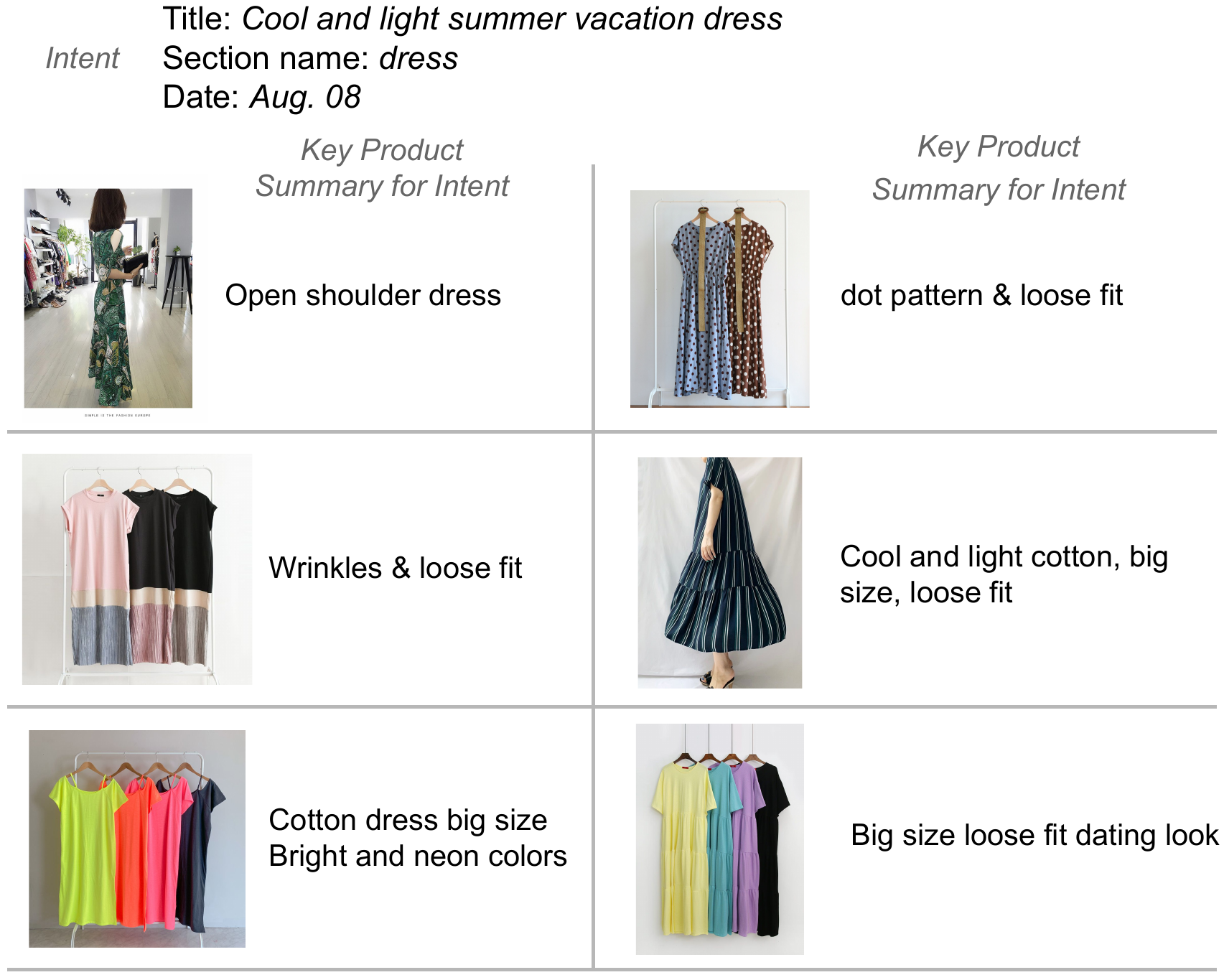}
\end{center}

\section{Example of product collection created using our models exposed on the service.}
\label{screenshot}
~\\[-20pt]
\begin{center}
\centering
\includegraphics[width=.73\linewidth]{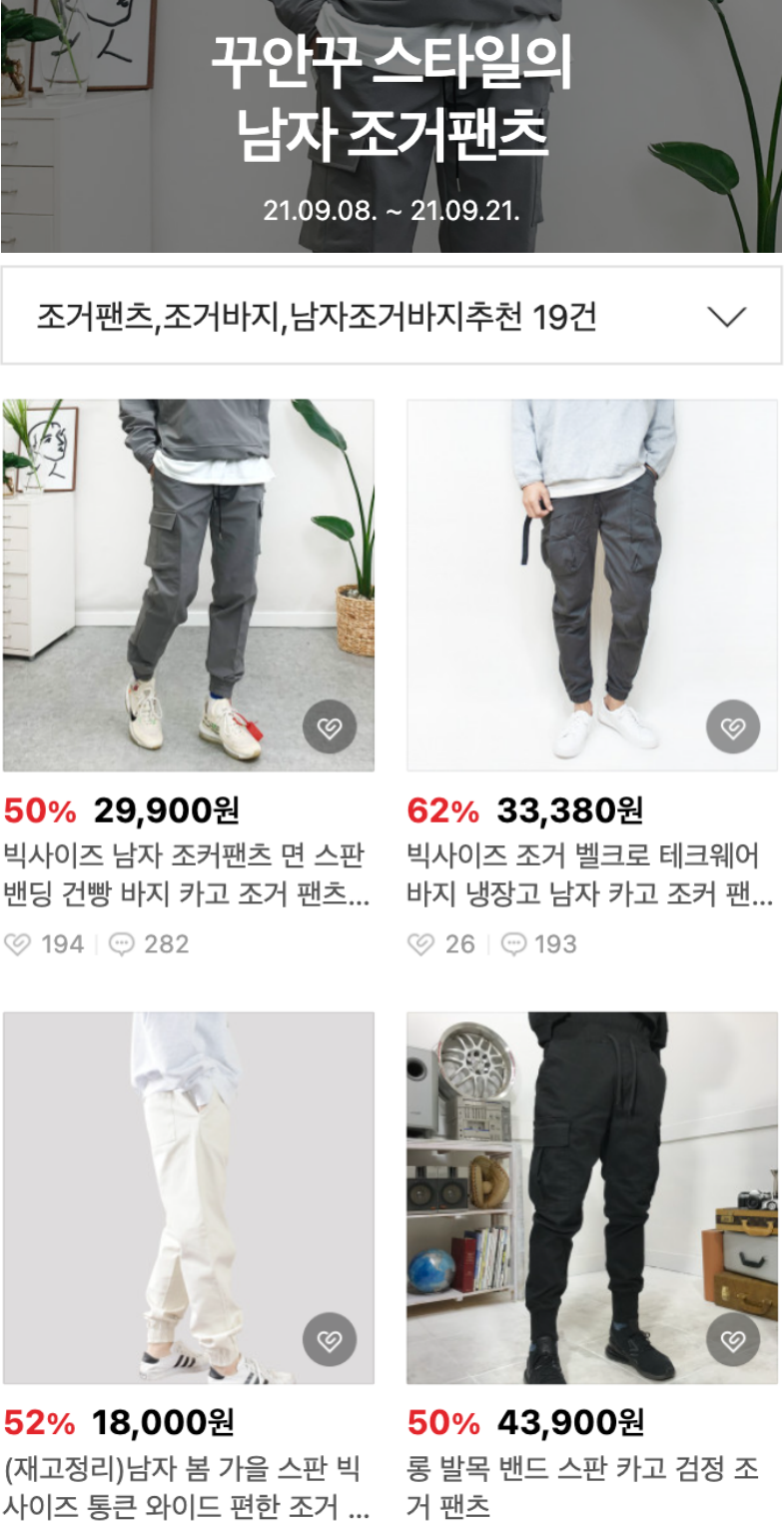}
~\\[3pt]
\end{center}
The title is ``kkuankku style men's jogger pants'' in English. Here, `kkuankku' is a Korean abbreviation that means `effortlessly chic'. Note that the title is generated by Korean GPT-3 HyperCLOVA~\cite{kim2021changes}.

\end{document}